\documentclass[aps,twocolumn,a4paper,pra,superscriptaddress]{revtex4}%
\usepackage{amssymb}
\usepackage{amsmath}
\usepackage{amsfonts}
\usepackage{graphicx}
\usepackage{graphics}

\def\Cas{\mathrm{Cas}}

\def\B{\mathrm{B}}

\def\bk{\mathbf{k}}

\def\dd{\mathrm{d}}

\def\TE{\mathrm{TE}}
\def\TM{\mathrm{TM}}

\def\P{\mathrm{P}}    
\def\C{\mathrm{C}}    
\def\lat{\mathrm{lat}}
\def\GC{G_\C} 
\def\rC{\rho_\C} 
\def\lat{\mathrm{lat}}  
\def\kC{k_\C}    
    
\def\max{\mathrm{max}}    
\def\min{\mathrm{min}}    
\def\perf{\mathrm{perf}}    
\def\Gold{\mathrm{Gold}}    
\def\PFA{\mathrm{PFA}}
\def\Tr{\mathrm{Tr}}
\def\calS{\mathcal{S}}
\def\calR{\mathcal{R}}
\def\calK{\mathcal{K}}
\def\calD{\mathcal{D}}
\def\calF{\mathcal{F}}
\def\calE{\mathcal{E}}

\begin{document}

\title{THE SCATTERING APPROACH TO THE CASIMIR FORCE}

\author{S. REYNAUD$^*$, A. CANAGUIER-DURAND, R. MESSINA, A. LAMBRECHT}
\address{Laboratoire Kastler Brossel, ENS, UPMC, CNRS, Jussieu, 75252 Paris, France\\
$^*$E-mail: serge.reynaud@upmc.fr}
\author{and P.A. MAIA NETO}
\address{Instituto de F\'{\i}sica, UFRJ, CP 68528, Rio de Janeiro,  RJ, 21941-972, Brazil}

\begin{abstract}
We present the scattering approach which is nowadays the best tool
for describing the Casimir force in realistic experimental configurations. 
After reminders on the simple geometries of 1d space and specular scatterers in 3d space,
we discuss the case of stationary arbitrarily shaped mirrors in electromagnetic vacuum.
We then review specific calculations based on the scattering approach,
dealing for example with the forces or torques between nanostructured surfaces 
and with the  force between a plane and a sphere.
In these various cases, we account for the material dependence of the forces,
and show that the geometry dependence goes beyond the trivial 
{\it Proximity Force Approximation} often used for discussing experiments.
\end{abstract}


\maketitle

\section*{The many facets of the Casimir effect}

The Casimir effect \cite{Casimir} is a jewel with many facets.
First, it is an observable effect of vacuum fluctuations in the mesoscopic world, 
which deserves careful attention as a crucial prediction of quantum field theory 
\cite{Milonni94,LamoreauxResource99,Bordag01,DeccaAP05,Milton05,LambrechtNJP06}.

Then, it is also a fascinating interface between quantum field theory and 
other important aspects of fundamental physics.
It has connections with the puzzles of gravitational physics through
the problem of vacuum energy \cite{GenetDark02,Jaekel08}
as well as with the principle of relativity of motion through
the dynamical Casimir-like effects \cite{Jaekel97,Lambrecht05,Braggio09}.
Effects beyond the Proximity Force Approximation also make apparent 
the extremely rich interplay of vacuum energy with geometry 
(references and more discussions below).

Casimir physics also plays an important role in the tests of gravity 
at sub-millimeter ranges \cite{Fischbach98,Adelberger03}.
Strong constraints have been obtained in short range
Cavendish-like experiments \cite{Kapner07} :
Should an hypothetical new force have a Yukawa-like form, its strength
could not be larger than that of gravity if the range is larger than 56$\mu$m.
For scales of the order of the micrometer, similar tests are performed
by comparing with theory the results of Casimir force measurements
\cite{LambrechtPoincare,OnofrioNJP06}. At even shorter scales,
the same can be done with atomic \cite{Lepoutre09} or nuclear 
\cite{Nesvizhevsky08} force measurements. 

Finally, the Casimir force and closely related Van der Waals force
are dominant at micron or sub-micron distances, which entails that 
they have strong connections with various important domains,
such as atomic and molecular physics, condensed matter and surface physics,
chemical and biological physics, micro- and nano-technology  
\cite{Parsegian06}.

\section*{Comparison of the Casimir force measurements with theory}

In short-range gravity tests, the new 
force would appear as a difference between the experimental result 
$F_\mathrm{exp}$ and theoretical prediction $F_\mathrm{th}$. 
This implies that $F_\mathrm{th}$ and $F_\mathrm{exp}$ 
have to be assessed independently from each other and
should forbid anyone to use theory-experiment comparison 
for proving (or disproving) some specific experimental result or theoretical model.

Casimir calculated the force between a pair of perfectly smooth, flat and
parallel plates in the limit of zero temperature and perfect reflection.
He found universal expressions for the force $F_\Cas$ and energy $E_\Cas$ 
\begin{eqnarray}
F_\Cas=\frac{\hbar c \pi ^2 A}{240L^4} \quad,\quad 
E_\Cas= - \frac{\hbar c \pi^2 A}{720 L^3} 
\end{eqnarray}
with $L$ the distance, $A$ the area, 
$c$ the speed of light and $\hbar$ the Planck constant.
This universality is explained by the saturation of the optical response of 
perfect mirrors which reflect 100\% (no less, no more) of the incoming fields. 
Clearly, this idealization does not correspond to any real mirror. 
In fact, the effect of imperfect reflection is large in most experiments, and
a precise knowledge of its frequency dependence is essential for obtaining a
reliable theoretical prediction for the Casimir force \cite{LambrechtEPJ00}.

The most precise experiments are performed with metallic mirrors which are 
good reflectors only at frequencies smaller than their plasma frequency $\omega_\P$.
Their optical response is described by a reduced dielectric function usually
written at imaginary frequencies $\omega=i\xi$ as 
\begin{eqnarray}
\varepsilon \left[i\xi\right] = \hat{\varepsilon}\left[i\xi\right] +
\frac{\sigma \left[i\xi\right] }{\xi} \quad,\quad 
\sigma \left[i\xi\right] = \frac{\omega_\P^2}{\xi+\gamma}
\end{eqnarray}
The function $\hat{\varepsilon} \left[i\xi\right] $ represents the contribution 
of interband transitions and it is regular at the limit $\xi\to0$. 
Meanwhile $\sigma \left[i\xi\right]$
is the reduced conductivity ($\sigma$ is measured as a frequency
and the SI conductivity is $\epsilon_0\sigma$)
which describes the contribution of the conduction electrons.

A simplified description corresponds to the lossless limit $\gamma \to 0$ 
often called the plasma model. As $\gamma$ is much smaller than
$\omega_\P$ for a metal such as Gold, this simple model captures
the main effect of imperfect reflection. However it cannot be considered 
as an accurate description since a much better fit of tabulated optical data 
is obtained with a non null value of $\gamma$ \cite{LambrechtEPJ00}.
Furthermore, the Drude model meets the important property of ordinary metals
which have a finite static conductivity $\sigma_0 = \frac{\omega_\P^2}{\gamma}$, 
in contrast to the lossless limit which corresponds to an infinite value 
for $\sigma_0$.

Another correction to the Casimir expressions is 
associated with the effect of thermal fluctuations \cite{Mehra67,Schwinger78} 
which is correlated to the effect of imperfect reflection \cite{GenetPRA00}. 
Bostrom and Sernelius have remarked that the small non zero value of $\gamma$
had a significant effect on the force evaluation at $T\neq0$ \cite{Bostrom00}. 
This remark has led to a blossoming of contradictory papers 
(see references in \cite{Reynaud03,BrevikNJP06,IngoldPRE09}).
The current status of Casimir experiments appears to favor predictions 
obtained with $\gamma=0$ rather than those corresponding to the 
expected $\gamma\neq0$ (see Fig.1 in \cite{DeccaPRD07}). Note that
the ratio between the prediction at $\gamma=0$ with that at $\gamma\neq0$ 
reaches a factor 2 at the limit of large temperatures or large distances,
although it is not possible to test this striking prediction with current 
experiments which do not explore this domain.

At this point, it is worth emphasizing that microscopic descriptions
of the Casimir interaction between two metallic bulks lead to predictions
agreeing with the lossy Drude model rather than the lossless plasma model
at the limit of large temperatures or large distances
\cite{Jancovici05,Buenzli05,Bimonte09}.
At the end of this discussion, we thus have to face a worrying situation
with a lasting discrepancy between theory and experiment.
This discrepancy may have various origins, in particular
artefacts in the experiments or inaccuracies in the calculations.
A more subtle but maybe more probable possibility is that there exist
yet unmastered differences between the situations studied in theory 
and the experimental realizations.

\section*{The role of geometry}

The geometry of Casimir experiments might play an important role in this context.
Precise experiments are indeed performed between a plane and a sphere
whereas calculations are often devoted to the geometry of two parallel planes.
The estimation of the force in the plane-sphere geometry involves the so-called 
\textit{Proximity Force Approximation} (PFA) \cite{Derjaguin68} which amounts to 
averaging over the distribution of local inter-plate distances the force 
calculated in the two-planes geometry, the latter being deduced 
from the Lifshitz formula \cite{Lifshitz56,DLP61}.

This trivial treatment of geometry cannot reproduce the rich interconnection
expected to take place between the Casimir effect and geometry \cite{Balian}.
In the plane-sphere geometry in particular, the PFA can only be valid when the 
radius $R$ is much larger than the separation $L$ \cite{Jaffe04}.
But even if this limit is met in experiments, the PFA does not tell one
what is its accuracy for a given value of $L/R$ or whether this accuracy
depends on the material properties of the mirror.
Answers to these questions can only be obtained by pushing the theory beyond 
the PFA, which has been done in the past few years (see references in 
\cite{ReynaudJPA08,EmigJPA08,BordagJPA08,WirzbaJPA08,KlingmullerJPA08}).
In fact, it is only very recently that these calculations have been done
with plane and spherical metallic plates coupled to electromagnetic vacuum 
\cite{CanaguierPRL09}, thus opening the way to a comparison with experimental
studies of PFA in the plane-sphere geometry \cite{Krause07}.

Another specific geometry of great interest is that of surfaces with periodic 
corrugations. As lateral translation symmetry is broken, the Casimir force 
contains a lateral component which is smaller than the normal one, but has
nevertheless been measured in dedicated experiments \cite{Chen02}.
Calculations beyond the PFA have first been performed with the simplifying
assumptions of perfect reflection \cite{Emig05} or shallow corrugations 
\cite{Maia05,Rodrigues06,Rodrigues07}.
As expected, the PFA was found to be accurate only at the limit
of large corrugation wavelengths. 
Very recently, experiments have been able to probe the beyond-PFA regime
\cite{Chan08,Chiu09} and it also became possible to calculate the forces
between real mirrors with deep corrugations \cite{Lambrecht08}. 
More discussions on these topics will be presented below.

\section*{Introduction to the scattering approach}

The best tool available for addressing these questions is the scattering
approach. We begin the review of this approach by an introduction
considering the two simple cases of the Casimir force between 2 scatterers 
on a 1-dimensional line and between two plane and parallel mirrors coupled
through specular scattering to 3-dimensional electromagnetic fields 
\cite{Jaekel91}.

The first case corresponds to the quantum theory of a scalar field with 
two counterpropagating components.
A mirror is thus described by a 2x2 $S-$matrix containing
the reflection and transmission amplitudes $r$ and $t$.
Two mirrors form a Fabry-Perot cavity described by a global $S-$matrix 
which can be evaluated from the elementary matrices $S_1$ and $S_2$
associated with the two mirrors. All $S-$matrices are unitary
and their determinants are shown to obey the simple relation
\begin{eqnarray}
&&ln\det S =  \ln\det S_1 + \ln\det S_2 + i \Delta \\
&&\Delta=i\ln\frac d{d^*} \quad,\quad
d (\omega) =1-r_1r_2\exp\left(\frac{2i\omega L}c \right) \nonumber
\end{eqnarray}
The phaseshift $\Delta$ associated with the cavity is expressed
in terms of the denominator $d$ describing the resonance effect. 
The sum of all these phaseshifts over the field modes leads to the 
following expression of the Casimir free energy $\calF$ 
\begin{eqnarray}
&&\calF = 
- \hbar\int\frac{\dd\omega}{2\pi} N(\omega) \Delta(\omega) \\
&&
N(\omega) = \frac1{\exp\left(\frac{\hbar\omega}{k_\B T}\right)-1} + \frac12 
\nonumber
\end{eqnarray}
Here $N$ is the mean number of thermal photons per mode,
given by the Planck law, augmented by the term $\frac12 $
which represents the contribution of vacuum \cite{Jaekel91}.

This phaseshift formula can be given alternative interpretations. 
In particular, the Casimir force 
\begin{eqnarray}
F=\frac{\partial\calF(L,T)}{\partial L} 
\end{eqnarray}
can be seen as resulting from the difference of radiation pressures exerted 
onto the inner and outer sides of the mirrors by the field fluctuations
\cite{Jaekel91}. 
Using the analytic properties of the scattering amplitudes, the free energy 
may be written as the following expression after a Wick rotation
($\omega=i\xi$ are imaginary frequencies)
\begin{eqnarray}
\mathcal{F} = 
\hbar\int\frac{\dd\xi}{2\pi} \cot\left(\frac{\hbar\xi}{2k_\B T}\right) \ln d(i\xi) 
\end{eqnarray}
Using the pole decomposition of the cotangent function and the analytic
properties of $\ln d$, this can finally be expressed as the Matsubara sum 
($\sum_m^\prime$ is the sum over positive integers $m$
with $m=0$ counted with a weight $\frac12$)
\begin{eqnarray}
&&\calF = k_\B T \sum_m{}^\prime \ln d(i\xi_m) 
\quad,\quad \xi_m \equiv \frac{2\pi m k_\B T}\hbar 
\end{eqnarray}

The same lines of reasoning can be followed when studying the geometry of 
two plane and parallel mirrors aligned along the axis $x$ and $y$.
Due to the symmetry of this configuration,
the frequency $\omega$, transverse vector $\bk \equiv \left( k_x,k_y\right)$ and
polarization $p=\TE,\TM$ are preserved by all scattering processes.
The two mirrors are described by reflection and transmission amplitudes 
which depend on frequency, incidence angle and polarization $p$.
We assume thermal equilibrium for the whole ``cavity + fields'' system, 
and calculate as in the simpler case of a 1-dimensional space.
Care has however to be taken to account for the contribution of evanescent waves 
besides that of ordinary modes freely propagating outside and inside the cavity 
\cite{GenetPRA03,LambrechtNJP06}. The properties of the evanescent waves are described 
through an analytical continuation of those of ordinary ones, using the well 
defined analytic behavior of the scattering amplitudes.
At the end of this derivation, the free energy 
has the following form as a Matsubara sum \cite{Matsubara}
\begin{eqnarray}
\label{CasimirFreeEnergy}
&&\calF = 
\sum_\bk \sum_p k_\B T \sum_m{}^\prime \ln d(i\xi_m,\bk,p) \\
&&d(i\xi,\bk,p) = 1 - r_1(i\xi,\bk,p) r_2(i\xi,\bk,p) 
\exp^{ -2\varkappa L } \nonumber \\
&&\xi_m \equiv \frac{2\pi m k_\B T}\hbar \quad,\quad 
\varkappa \equiv \sqrt{\bk^2+\frac{\xi^2}{c^2}} \nonumber 
\end{eqnarray}
$\sum_\bk \equiv A \int\frac{\dd^2\bk}{4\pi^2}$ is
the sum over transverse wavevectors with $A$ the area of the plates, 
$\sum_p$ the sum over polarizations and $\sum_m{}^\prime$ the 
Matsubara sum.

This expression reproduces the Casimir ideal formula
in the limits of perfect reflection $r_1 r_2 \rightarrow 1$
and null temperature $T \rightarrow 0$. 
But it is valid and regular at thermal equilibrium at any temperature
and for any optical model of mirrors obeying causality and high frequency 
transparency properties.
It has been demonstrated with an increasing range of validity
in \cite{Jaekel91}, \cite{GenetPRA03} and \cite{LambrechtNJP06}. 
The expression is valid not only for lossless mirrors but also 
for lossy ones. In the latter case, it accounts for the additional
fluctuations accompanying losses inside the mirrors.

It can thus be used for calculating the Casimir force between arbitrary
mirrors, as soon as the reflection amplitudes are specified. 
These amplitudes are commonly deduced from models of mirrors,
the simplest of which is the well known Lifshitz model 
\cite{Lifshitz56,DLP61} which
corresponds to semi-infinite bulk mirrors characterized by a 
local dielectric response function $\varepsilon (\omega)$
and reflection amplitudes deduced from the Fresnel law.

In the most general case, the optical response of the mirrors
cannot be described by a local dielectric response function.
The expression (\ref{CasimirFreeEnergy}) of the free energy is still valid
in this case with some reflection amplitudes to be determined from 
microscopic models of mirrors. 
Recent attempts in this direction can be found for example in
\cite{Pitaevskii08,Dalvit08,Svetovoy08}. 

\section*{The non-specular scattering formula}

We now present a more general scattering formula allowing one to calculate 
the Casimir force between stationary objects with arbitrary non planar shapes.
The main generalization with respect to the already discussed cases is that
the scattering matrix $\calS$ is now a larger matrix accounting for 
non-specular reflection and mixing different wavevectors and polarizations
while preserving frequency \cite{Maia05,LambrechtNJP06}.
Of course, the non-specular scattering formula is the generic one
while specular reflection can only be an idealization.

As previously, the Casimir free energy can be written as the sum 
of all the phaseshifts contained in the scattering matrix $\calS$
\begin{eqnarray}
\calF &=& i\hbar \int_0^\infty \frac{\dd \omega}{2\pi}
N(\omega) \ln \det \calS \nonumber \\
&=&i\hbar \int_0^\infty \frac{\dd \omega}{2\pi}
N(\omega) \Tr \ln \calS 
\end{eqnarray}
The symbols $\det$ and $\Tr$ refer to determinant and trace over the modes of
the matrix $\calS$.
As previously, the formula can also be written after a Wick rotation
as a Matsubara sum
\begin{eqnarray}
\label{CasimirFreeEnergyNS}
&&\calF = k_\B T \sum_m{}^\prime \Tr \ln \calD (i\xi_m) \\ 
&&\calD = 1 - \calR_1 \exp^{ -\calK L } \calR_2 \exp^{ -\calK L } \nonumber
\end{eqnarray}
The matrix $\calD$ is the denominator containing all the resonance 
properties of the cavity formed by the two objects 1 and 2
here written for imaginary frequencies.
It is expressed in terms of the matrices $\calR_1$ and $\calR_2$ which represent 
reflection on the two objects 1 and 2 and of propagation factors $\exp^{-\calK L}$. 
Note that the matrices $\calD$, $\calR_1$ and $\calR_2$, which were diagonal 
on the basis of plane waves when they described specular scattering,  
are no longer diagonal in the general case of non specular scattering. 
The propagation factors remain diagonal in this basis with their diagonal values 
written as in (\ref{CasimirFreeEnergy}). Clearly the expression 
(\ref{CasimirFreeEnergyNS}) does not depend on the choice of a specific basis. 
Remark also that (\ref{CasimirFreeEnergyNS}) takes a simpler form at the limit 
of null temperature (note the change of notation from the free energy $\calF$ 
to the ordinary energy $\calE$)
\begin{eqnarray}
&&F=\frac{\dd \calE}{\dd L} \quad,\quad
\calE = \hbar \int_0^\infty \frac{\dd \xi}{2\pi} \ln \det \calD (i\xi) 
\label{CasimirEnergyTnull}
\end{eqnarray}

Formula (\ref{CasimirEnergyTnull}) has been used to evaluate the effect of roughness 
or corrugation of the mirrors \cite{Maia05,Rodrigues06,Rodrigues07} in a
perturbative manner with respect to the roughness or corrugation amplitudes
(see the next section).
It has clearly a larger domain of applicability, not limited to the 
perturbative regime, as soon as techniques are available for computing 
the large matrices involved in its evaluation.
It has also been used in the past years by different groups using different notations
\cite{EmigJPA08,BordagJPA08,WirzbaJPA08,Emig08,Kenneth08}.
The relation between these approaches is reviewed 
for example in \cite{Milton08}.

\section*{The lateral Casimir force between corrugated plates}

As already stated, the lateral Casimir force between corrugated plates
is a topic of particular interest. This configuration is more
favorable to theory/experiment comparison than that met when 
studying the normal Casimir force. It could thus allow for a new 
test of Quantum ElectroDynamics, through the dependence of
the lateral force to the corrugation wavevector 
\cite{Rodrigues06,Rodrigues07}.
Here, we consider two plane mirrors, M1 and M2, with corrugated surfaces
described by uniaxial sinusoidal profiles (see Fig.~1 in \cite{Rodrigues07}).
We denote $h_1$ and $h_2$ the local heights with respect to mean planes 
$z_1=0$ and $z_2=L$
\begin{eqnarray}
h_1=a_1\,\cos(\kC x)\quad,\quad h_2=a_2\,\cos\left(\kC (x-b)\right)
\end{eqnarray}
$h_1$ and $h_2$ have null spatial averages and $L$ is the mean distance 
between the two surfaces;
$h_1$ and $h_2$ are both counted as positive when they correspond to 
separation decreases;
$\lambda_\C$ is the corrugation wavelength, 
$\kC=2\pi/\lambda_\C$ the corresponding wave vector,
and $b$ the spatial mismatch between the corrugation crests.

At lowest order in the corrugation amplitudes, when 
$a_1, a_2 \ll \lambda_\C, \lambda_\P, L$,
the Casimir energy may be obtained by expanding up to second order 
the general formula (\ref{CasimirEnergyTnull}).
The part of the Casimir energy able to produce a lateral force 
is thus found to be 
\begin{eqnarray}
&&F^\lat = -\frac{\partial \delta \calE}{\partial b} \\
&& \delta \calE = 
- \hbar \int_0^{\infty} \frac{\dd\xi}{2\pi}
\Tr \left( \delta\calR_1 \frac{\exp^{-\calK L}}{\calD_0} 
\delta\calR_2 \frac{\exp^{-\calK L}}{\calD_0}  \right) \nonumber
\label{Ecorrug}
\end{eqnarray}
$\delta\calR_1$ and $\delta\calR_2$ are the first-order variation
of the reflection matrices $\calR_1$ and $\calR_2$ induced by the 
corrugations; 
$\calD_0$ is the matrix $\calD$ evaluated at zeroth order in the
corrugation; it is diagonal on the basis of plane waves and 
commutes with $\calK$.

Explicit calculations of (\ref{Ecorrug}) have been done for the simplest
case of experimental interest, with two corrugated metallic plates 
described by the plasma dielectric function.
These calculations have led to the following expression of the lateral energy
\begin{eqnarray}
\delta E = \frac A2 \GC(\kC) a_1 a_2 \cos (\kC b) 
\label{spectral}
\end{eqnarray}
with the function $\GC(\kC)$ given in \cite{Rodrigues07}.
It has also been shown that the PFA was recovered for long corrugation wavelengths,
when $\GC(\kC)$ is replaced by $\GC(0)$ in (\ref{spectral}).
This important argument can be considered as a properly formulated 
``Proximity Force Theorem'' \cite{Rodrigues07}.
It has to be distinguished from the approximation (PFA)
which consists in an identification of $\GC(\kC)$ with its limit $\GC(0)$.
For arbitrary corrugation wavevectors, the deviation from the PFA is 
described by the ratio 
\begin{eqnarray}
\rC(\kC)=\frac{\GC(\kC)}{\GC(0)}
\end{eqnarray}
The variation of this ratio $\rC$ with the parameters of interest
has been described in a detailed manner in \cite{Rodrigues06,Rodrigues07}.
Curves are drawn as examples in the Fig.~1 of \cite{Rodrigues06}
with $\lambda_\P=137$nm chosen to fit the case of gold covered plates. 
An important feature is that $\rC$ is smaller than unity as soon as
$\kC$ significantly deviates from 0.
For large values of $\kC$, it even decays exponentially to zero,
leading to an extreme deviation from the PFA.

Other situations of interest have also been studied.
When the corrugation plates are rotated with respect to each other,
a torque appears to be induced by vacuum fluctuations, tending to
align the corrugation directions \cite{RodriguesEPL06}.
In contrast with the similar torque appearing between misaligned
birefringent plates \cite{Munday05}, the torque is here coupled to
the lateral force. The advantage of the configuration with
corrugated plates is that the torque has a larger magnitude.
Another case of interest may be designed by using the possibilities
offered by cold atoms techniques. Non trivial effects of geometry
should be visible in particular when using a Bose-Einstein condensate
as a local probe of vacuum above a nano-grooved plate
\cite{Dalvit08a,Messina09}.

These results suggested that non trivial effects of geometry, \textit{i.e.}
effects beyond the PFA, could be observed with dedicated lateral force experiments.
It was however difficult to achieve this goal with corrugation amplitudes $a_1, a_2$ 
meeting the conditions of validity of the perturbative expansion.
As already stated, recent experiments have been able to probe the beyond-PFA regime
with deep corrugations \cite{Chan08,Chiu09} and it also became possible to calculate 
the forces between real mirrors without the perturbative assumption.
In particular, an exact expression has been obtained for the force between
two nanostructured surfaces made of real materials with arbitrary 
corrugation depth, corrugation width and distance \cite{Lambrecht08}. 

\section*{The plane-sphere geometry beyond PFA}

In the plane-sphere geometry, it is also possible to use the general scattering 
formula (\ref{CasimirEnergyTnull}) to obtain explicit evaluations of the Casimir force.
The reflection matrices may here be written in terms of Fresnel amplitudes
on the plane mirror and of Mie amplitudes on the spherical one.
The scattering formula is then obtained by writing also transformation formulas
from the plane waves basis to the spherical waves basis and conversely.
The result takes the form of a multipolar expansion with spherical waves
labeled by quantum numbers $\ell$ and $m$ ($\vert m\vert\le\ell$).
For doing the numerics, the expansion is truncated at some maximum value $\ell_\max$,
which restricts accurate evaluations to a domain $x\equiv L/R>x_\min$ with
$x_\min$ proportional to $1/\ell_\max$. 

Such calculations have first been performed for perfectly reflecting mirrors
\cite{Maia08,Emig08}. It was thus found that the Casimir energy was smaller 
than expected from the PFA and, furthermore, than the result for 
electromagnetic fields was departing from PFA more rapidly than was expected
from previously existing scalar calculations \cite{WirzbaJPA08,BordagJPA08}.
It is only very recently that the same calculations have been done for the
more realistic case of metallic mirrors described by a plasma model dielectric
function \cite{CanaguierPRL09}. Results of these evaluations are expressed 
in terms of reduction factors defined for the force $F$ or force gradient $G$ 
with respect to the PFA expectations $F^\PFA$ and $G^\PFA$ respectively 
\begin{eqnarray}
\rho_F=\frac{F}{F^\PFA} \quad,\quad \rho_G=\frac{G}{G^\PFA}
\end{eqnarray}
Examples of curves for $\rho_F$ and $\rho_G$ are shown on Fig.2 
of \cite{CanaguierPRL09} for perfect and plasma mirrors.

Using these results, it is possible to compare the theoretical evaluations
to the experimental study of PFA in the plane-sphere geometry \cite{Krause07}.
In this experiment, the force gradient is measured for various radii of the
sphere and the results are used to obtain a constraint $\vert\beta_G\vert<0.4$ 
on the slope at origin $\beta_G$ of the function 
\begin{eqnarray}
\rho_G(x)=1+\beta_G x+O(x^2)
\end{eqnarray}
Now the comparison of this experimental information to the slope obtained
by interpolating at low values of $x$ the theoretical evaluations of $\rho_G$
reveals a striking difference between the cases of perfect and plasma mirrors.
The slope $\beta_G^\perf$ obtained for perfect mirrors is larger than that 
$\beta_G^\Gold$ obtained for gold mirrors by a factor larger than 2
\begin{eqnarray}
\beta_G^\perf \sim-0.48 \quad,\quad \beta_G^\Gold \sim-0.21 
\end{eqnarray}
Meanwhile, $\beta_G^\Gold$ is compatible with the experimental bound 
obtained in \cite{Krause07} (see \cite{CanaguierPRL09}) 
whereas $\beta_G^\perf$ lies outside this bound
(see also \cite{Emig08}).

The lesson to be learned from these results is that more
work is needed to reach a reliable comparison of experiment and theory
on the Casimir effect. Experiments are performed with large spheres
for which the parameter $L/R$ is smaller than 0.01, and efforts are
devoted to calculations pushed towards this regime \cite{Bordag09}.

Meanwhile, the effect of temperature should also be correlated
with the plane-sphere geometry.
The first calculations accounting simultaneously for plane-sphere geometry,
temperature and dissipation have been published very recently \cite{Canaguier10}
and they show several striking features. 
The factor of 2 between the long distance forces in Drude and plasma models 
is reduced to a factor below 3/2 in the plane-sphere geometry.
Then, PFA underestimates the Casimir force within the Drude model at short distances, 
while it overestimates it at all distances for the perfect reflector and plasma model.
If the latter feature were conserved for the experimental parameter region $R/L$ $(>10^2)$, 
the actual values of the Casimir force calculated within plasma and Drude model could turn 
out to be closer than what PFA suggests, which would diminish the discrepancy between 
experimental results and predictions of the thermal Casimir force using the Drude model.

\section*{Acknowledgments}
The authors thank I. Cavero-Pelaez, D. Dalvit, G.L. Ingold, M.-T. Jaekel 
and I. Pirozenkho for fruitful discussions. 
A.C. and R.M. acknowledge support from the ESF Research Networking Programme 
CASIMIR (www.casimir-network.com).
P.A.M.N. thanks CNPq, CAPES and Faperj for financial support.
A.L. acknowledges support from the French Contract ANR-06-Nano-062.


\begin{thebibliography}{99}
\bibitem{Matsubara}
The question had been asked in \cite{Reynaud03} whether the regularity conditions
needed to write the Matsubara sum were met for the Drude model; this question
has been answered positively in \cite{IngoldPRE09}.
\end{thebibliography}
\end{document}